\documentclass[aps,pre,twocolumn,nofootinbib,tightenlines,superscriptaddress,nopacs,amsmath,amssymb,final,letterpaper]{revtex4-1}

\usepackage[utf8]{inputenc}
\usepackage{calc}
\usepackage{graphicx}
\usepackage{amsmath,amssymb,amsthm}
\usepackage{bbold}
\usepackage{bm}
\usepackage{color}
\usepackage[dvipsnames]{xcolor}
\usepackage{enumitem}
\usepackage{tikz}
\usepackage{float}
\usepackage[percent]{overpic}

\newcommand{\abs}[1]{\vert #1\vert}

\newcommand{\expec}[1]{\langle #1\rangle}

\def\multiset#1#2{\ensuremath{\left(\kern-.3em\left(\genfrac{}{}{0pt}{}{#1}{#2}\right)\kern-.3em\right)}}

\usepackage[colorlinks=true,linkcolor=blue,urlcolor=blue,citecolor=blue,anchorcolor=blue]{hyperref}

\begin{document}
\title{Identifying hubs in directed networks}

\author{Alec \surname{Kirkley}}
\email{alec.w.kirkley@gmail.com}
\affiliation{Institute of Data Science, University of Hong Kong, Hong Kong}
\affiliation{Department of Urban Planning and Design, University of Hong Kong, Hong Kong}
\affiliation{Urban Systems Institute, University of Hong Kong, Hong Kong}

\begin{abstract}
Nodes in networks that exhibit high connectivity, also called ``hubs'', play a critical role in determining the structural and functional properties of networked systems. However, there is no clear definition of what constitutes a hub node in a network, and the classification of network hubs in existing work has either been purely qualitative or relies on ad hoc criteria for thresholding continuous data that do not generalize well to networks with certain degree sequences. Here we develop a set of efficient nonparametric methods that classify hub nodes in directed networks using the Minimum Description Length principle, effectively providing a clear and principled definition for network hubs. We adapt our methods to both unweighted and weighted networks and demonstrate them in a range of example applications using real and synthetic network data. 
\end{abstract}


\maketitle

\section{Introduction}

Highly connected ``hub'' nodes play an important role in the structure and function of networks across a wide range of applications \cite{newman18c}. Hub regions in brain networks are central for communication and information integration \cite{van2013network,bassett2017network}. Hub stations in transportation networks are important for resilience to failures and efficient routing \cite{verma2014revealing,roucolle2020measuring}. Hub proteins in protein interaction networks are often essential for the survival and reproduction of an organism \cite{he2006hubs}. And hub locations in human mobility networks can be hotspots for congestion, economic activity, and the spread of disease \cite{mimar2022connecting,aguilar2022impact}.

The natural way to define a hub node in a network is to use degree centrality as an indicator---the more connections a node has, the more critical it is for network connectivity, and above some degree threshold (typically at or above the average degree in the network) we consider a node to be a ``hub'' \cite{Barabasi16}. Often the label of ``hub'' is reserved for nodes with an ``unusually high'' degree \cite{newman18c}, as these nodes have a disproportionate influence on many processes that take place on the network such as epidemics or information spreading \cite{chakrabarti2008epidemic, gradon2021countering}. The labelling of a node as a hub can be based on its in-degree and/or out-degree when the network is directed, depending on the application of interest. For example, in human mobility networks, one is often interested in targeting locations (nodes) with high population in-flows (weighted in-degree) for interventions to reduce congestion or the spread of disease, making the in-degree a relevant criterion for hub classification. 

The concept of a hub node can also be extended to capture more global notions of centrality in a network. For example, the HITS algorithm \cite{kleinberg1999authoritative} assigns a hub and authority score to each node in the network in a self-consistent manner: nodes are given a high authority score if they are pointed to by nodes with high hub scores, and nodes are given a high hub score if they point towards nodes with high authority scores. The stationary solution for the hub  scores in the HITS algorithm are the entries in the leading eigenvector of the adjacency matrix of the network multiplied by its transpose, which indicates that these hub scores are capturing global information about the graph. Any number of other global centrality indices such as closeness, betweenness, eigenvector centrality, or Katz centrality can also in principle be used to define hub nodes at a global level, but ultimately many of these measures are highly correlated with degree in a large number of network models and real-world systems \cite{oldham2019consistency,ronqui2015analyzing,grando2016analysis,schoch2017correlations,bloch2023centrality,shao2018rank}. 

Identifying hub nodes as nodes with ``unusually high'' degrees can be thought of as performing outlier detection on the degree sequence listing the degree of each node in the network. But existing information theoretic outlier detection methods either require the number of outliers to be known ahead of time \cite{wu2011information}, are formulated for general graph databases \cite{akoglu2015graph}, or have parametric forms for the model likelihood which must be inferred \cite{bohm2009coco}, making them poorly suited for direct application to network degree sequences for classifying hub nodes. 

The identification of hubs in weighted, directed networks also directly relates to the idea of identifying ``hotspots'' with high flows in human mobility networks, for which the ``Loubar'' method of \cite{louail2014mobile} is an elegant and widely used method. The Loubar method utilizes the Lorenz curve of the flows in and/or out of nodes in a human mobility network to identify nodes with high flows as hotspots of activity. The Loubar method has been used in range of applications to understand epidemic spreading \cite{hazarie2021interplay,aguilar2022impact}, commuting structure \cite{louail2015uncovering}, and economic growth \cite{xu2019inverted} using human mobility data. This method has the desirable property of being completely nonparametric---it automatically selects the number of hotspot nodes from the data itself, without any user-controlled input parameters. However, it does not specifically look at the pairwise nature of network structure in its formulation, so cannot be compared with other network models using rigorous model selection criteria. The Loubar method also does not depend on the full distribution of flows, but only the mean and maximum of the flows, which we show in a range of experiments can lead to undesirable behavior where a large fraction of nodes in the network are classified as hubs.  

Methods directly aimed at network compression using Bayesian or information theoretic criteria---including blockmodeling \cite{peixoto2019bayesian}, configuration models \cite{hebert2022network}, core-periphery modeling \cite{gallagher2021clarified}, and other nonparametric information theoretic methods for summarizing network structure \cite{kirkley2022spatial,kirkley2023compressing,kirkley2023constructing}---are well-suited for identifying hubs in network data, since these methods allow for the automatic selection of the number of hubs and comparison with other network models using the MDL principle. However, none of the aforementioned methods explicitly aim to identify hub nodes, and so may only identify hub nodes as a separate group by chance if grouping these nodes happens to provide compression with respect to the mixing structure assumed by the model. For example, high degree nodes are often grouped together in the (non-degree-corrected) stochastic blockmodel \cite{karrer2011stochastic}, and the core nodes in a core-periphery blockmodel are often of high degree \cite{gallagher2021clarified}. To identify hub nodes one only needs to consider the number of edges incident on the hub nodes, and not the connectivity among the hub nodes nor among the non-hub nodes which are key factors in the description length of a network under a blockmodel.   

In this paper we develop principled, efficient, nonparametric methods to classify hub nodes in directed networks based on the Minimum Description Length (MDL) principle, which states that the best model among a set of candidate models for a dataset is the one that results in the lowest description length for the data \cite{rissanen1978}. In our approach, the MDL principle allows us to select the optimal configuration of hub nodes in a network by minimizing the description length of data encodings that exploit degree discrepancies between hub nodes and non-hub nodes.  We adapt our formulation to multiple encoding schemes applicable to unweighted and weighted networks, and describe a simple, fast algorithm to identify the hub nodes in these networks. We apply our method to a variety of synthetic network models, finding that the extent to which we can compress networks with hub nodes depends on the heterogeneity of the degree distribution and that our encodings can give more intuitive summaries of the hub structure than existing methods in these cases. We also apply our method to growing random graph models, finding that we can identify a hub transition at which it becomes most compressive to describe the network using hub nodes, and that this transition depends on the parameters of the growth model in a physically meaningful way. Finally, we apply our method to a corpus of directed network datasets from a wide range of disciplines, finding that many real networks do not have hub structure according to our more conservative encoding and that the information in many of the networks with discernible hub structure can be effectively compressed by focusing on the high degree hub nodes when transmitting the network. 


\section{Methods}
\label{sec:methods}

To identify a particular type of structural and/or dynamical regularity in network data---for example, communities \cite{peixoto2019bayesian} or temporal change-points \cite{kirkley2023compressing}---one can first construct an information encoding that is designed to exploit this regularity given an input data classification---for example, a partition of the nodes into communities or a segmentation of a time series of networks. Then, given an appropriate encoding, one can minimize its description length over all data classifications to find a representation that succinctly describes the data by exploiting the desired property. This process is equivalent to Maximum A Posteriori (MAP) estimation with hierarchical Bayesian generative models \cite{peixoto2019bayesian} but can often provide a more intuitive problem framing when choosing among various data encodings (models). In this section we describe how to apply this line of reasoning for identifying hubs in network data.

\subsection{Compressing network data}
\label{sec:compression}

Let $G=(V,E)$ be a directed graph with $N$ nodes in the node set $V$ and $M$ edges in the edge set $E$. We will first treat the case where $G$ is a simple graph---in other words, $G$ has no edges from a node to itself and has at most one edge going from any node $i$ to any other node $j$. We will generalize our method to the multigraph case with self-edges in Sec.~\ref{sec:multigraph}. We will let $\bm{k}$ be the in-degree sequence such that $k_i$ is the in-degree of node $i$ and $\sum_{i=1}^{N}k_i=M$.

Now, suppose we aim to transmit the network $G$---or, equivalently, the source node $i\in V$ and target node $j\in V$ of all the edges $(i,j)\in E$---in binary to a receiver through some communication channel. We assume that the receiver knows $N$ and $M$, which would be of comparatively negligible cost to communicate and can be ignored anyway. Since there are $N(N-1)$ distinct ordered node pairs, and $M$ of these pairs contain an edge, then there are ${N(N-1)\choose M}$ possible graphs $G$, and the same number of possible binary messages we could end up transmitting to the receiver. $\lceil \log_2 {N(N-1)\choose M}\rceil$ bits will be enough to encode all such messages uniquely when establishing a codebook for our transmission ahead of time with the receiver, and so the information content or \emph{description length} of this na\"ive encoding of the graph $G$ is 
\begin{align}\label{eq:L0ERs}
\mathcal{L}^{(\text{ERs})}_0 = \log {N(N-1)\choose M}    
\end{align}
bits. Here we've ignored the ceiling function as it will provide a negligible change to the description length for $N \gg 1$, we've used the notation convention $\log_2 \equiv \log$, and we've used the subscript ``$0$'' to indicate that no hub nodes were used to aid in the network transmission process---we will describe how this works shortly. We use the superscript ``ERs'' to refer to the Erdos-Renyi model for simple directed graphs, $G(N,M)$, which selects uniformly at random from all directed simple graphs with $N$ nodes and $M$ edges. We can also derive Eq.~\ref{eq:L0ERs} as the negative log-probability of picking any particular graph $G$ from this ensemble. 

Alternatively, we can transmit the graph $G$ using a multi-part encoding, where each step involves utilizing a different codebook with the receiver that is constructed while keeping in mind the constraints imposed by information transmitted earlier in the transmission process. This process involves transmitting $G$ in increasing levels of granularity until the entire edge set $E$ is known. 

One very simple multi-part transmission scheme would involve first transmitting the in-degrees $\bm{k}=\{k_1,...,k_N\}$, then transmitting the entire edge set $E$ from the set of all edge sets consistent with the in-degrees $\bm{k}$. There are at most $\multiset{N}{M}={M+N-1\choose N-1}$ (where $\multiset{}{}$ is the multiset coefficient) unique ways to assign the $N$ nodes as targets in the $M$ directed edges---allowing nodes to potentially have in-degree $0$---in order to fully specify the in-degrees $\bm{k}$. This is the total number of messages that will be in the codebook for the first step of the process. (Depending on $M$ and $N$, not all of the ${M+N-1\choose N-1}$ in-degree sequences will correspond to valid directed simple graphs, so we are technically accounting for some ``non-graphical'' in-degree sequences in our encoding in addition to all valid in-degree sequences.) Then, once $\bm{k}$ is known, there are ${N-1\choose k_i}$ possible source nodes for the $k_i$ edges containing node $i$ as a target, resulting in $\prod_{i=1}^{N}{N-1\choose k_i}$ possible messages for the second step of the process. Adding the information content of these two steps, the description length of this alternative encoding is
\begin{align}\label{eq:L0CMs}
\mathcal{L}^{(\text{CMs})}_0 = \log {M+N-1\choose N-1} + \sum_{i=1}^{N}\log {N-1\choose k_i}  
\end{align}
bits. We use the superscript ``CMs'' to refer to the Configuration Model---with only in-degree constraints---for simple directed graphs, and again use a subscript ``$0$'' to indicate that no hub nodes were used to aid in the network transmission process. This model will select uniformly at random from all directed simple graphs with $N$ and $M$ edges and a specific in-degree sequence $\bm{k}$. We can alternatively derive Eq.~\ref{eq:L0CMs} as the negative log-probability of picking any particular graph $G$ from this ensemble, given a uniform prior over all degree sequences $\bm{k}$ of length $N$ that sum to $M$. 

One can show that for $N\gg\expec{k}\gg 1$ we will always achieve superior data compression---in other words, a lower description length for the network $G$---using the two-step encoding corresponding to  Eq.~\ref{eq:L0CMs} than the one-step encoding corresponding to Eq.~\ref{eq:L0ERs} (see Appendix~\ref{appendix:relative0ercm}). This suggests that multi-step encodings which exploit in-degree heterogeneity can provide improved network compression. 

\subsection{Compressing simple directed graphs with hub nodes}
\label{sec:simple}

\begin{figure*}
    \centering
    \includegraphics[width=\textwidth]{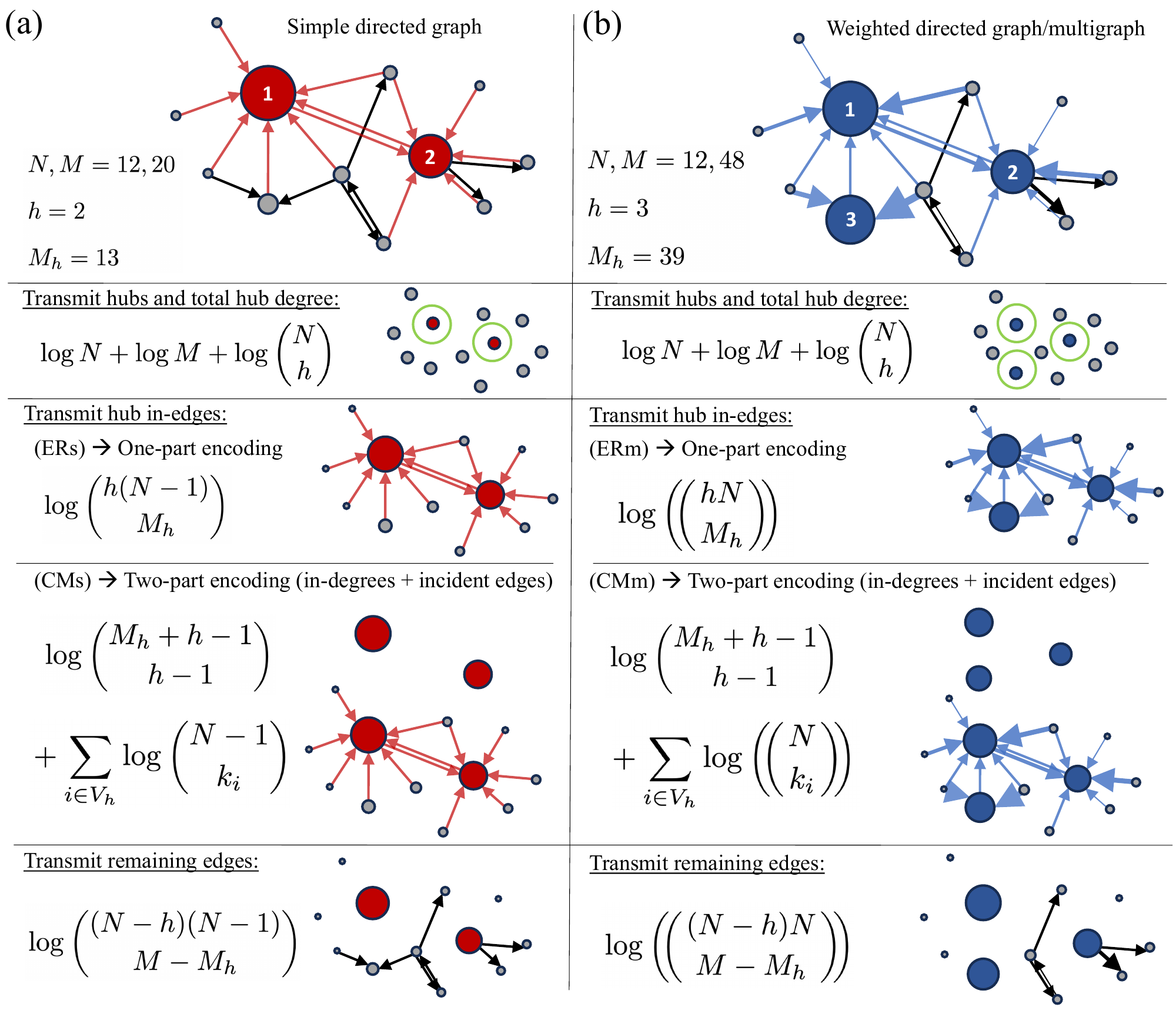}
    \caption{
    \textbf{Diagram of hub-based encodings.} (a) Schematic of the simple directed graph encoding described in Sec.~\ref{sec:simple}, along with the description length of each step. (b) Schematic of the weighted directed graph/multigraph encoding described in Sec.~\ref{sec:multigraph}, along with the description length of each step. We define the hub nodes of a network $G=(V,E)$ according to a given encoding (ERs, CMs, ERm, or CMm) as the node subset $V_h\subseteq V$ that minimizes the information required to transmit the network (e.g. the positions of the edges $E$) when transmitting the positions of edges incident to the hubs first. This provides a principled, nonparametric criterion for identifying hubs in directed networks based on the Minimum Description Length (MDL) principle.
    }
    \label{fig:diagram}
\end{figure*}

We can consider going one step further and transmitting the edges incident to a small set of high in-degree nodes---the ``hub'' nodes in the encoding---independently from the rest of the edges in the network. This will provide good compression when a large portion of the edges in $E$ are pointing towards a small set of hub nodes, because in this case there are comparatively few potential configurations of edges incident to the hubs---there are not many target nodes to choose from---and the total number of binary messages needed for encoding the specific edge configuration $E$ is reduced.

Consider the case where we have $h$ hub nodes $V_h\subseteq V$ and $N-h$ non-hub nodes $V\setminus V_h$, and we wish to transmit the edges incident to these two node sets independently to achieve improved data compression for the total edge set $E$. (We will address the issue of identifying the optimal value of $h$ later on.) Let $E_h$ denote the set of edges incident to the hub nodes $V_h$ as targets, and $M_h=\abs{E_h}$ be the number of such edges. Since there are $N$ possibilities for the value of $h$ and $M$ possibilities for the value of $M_h$, we will need $\log N+\log M=\log NM$ bits of information to transmit these initial quantities. 

Next we need to transmit the identities of the $h$ hub nodes $V_h$, which will require $\log{N\choose h}$ bits of information since it requires specifying a subset of $h$ nodes from the total set of $N$ nodes. Then, we can perform two transmission steps reminiscent of the one used to derive Eq.~\ref{eq:L0ERs}---one for the hub-incident edges $E_h$, and one for the rest of the edges $E\setminus E_h$. Specifying $E_h$ will require us to specify $M_h$ edge positions out of $h(N-1)$ total (ordered) node pairs, and specifying $E\setminus E_h$ will require us to specify $M-M_h$ edge positions out of the $(N-h)(N-1)$ remaining node pairs. The total information cost of this encoding is then given by
\begin{align}\label{eq:LERs}
\mathcal{L}^{(\text{ERs})}(V_h) &= \log NM + \log {N\choose h}  \\
&~+ \log {h(N-1)\choose M_h} + \log {(N-h)(N-1)\choose M-M_h} \nonumber,   
\end{align}
where we've used the ``ERs'' notation as before to denote the use of an encoding that transmits the edge positions for each node set in one step, as in Eq.~\ref{eq:L0CMs}. We've also made explicit that this description length of our hub-based encoding is a function of the number of hubs $h$ that we choose and which $h$ nodes $V_h$ we choose to be the hubs. For the cases $h=0$ and $h=N$, we let $\mathcal{L}^{(\text{ERs})}=\mathcal{L}^{(\text{ERs})}_0$, since the initial transmission costs are no longer needed.

We can view Eq.~\ref{eq:LERs} as an objective function that, when minimized over node subsets $V_h$, finds an optimal set of nodes to classify as hubs in our network. In other words, the optimal subset of nodes to identify as hubs is the subset of nodes that allows us to most parsimoniously describe (i.e. best compress) the network structure using an encoding aimed at exploiting structural heterogeneity between hubs and non-hubs. In Appendix~\ref{appendix:opt-ER} we show that the optimal choice for the hub nodes at any given value of $h$ is the set of nodes that maximizes $M_h$---in other words, the nodes with the $h$ highest in-degrees---which confirms the intuition behind the construction of the hub-based encoding. This implies that we can identify the globally optimal configuration of hub nodes in a network $G$ using the following simple algorithm:
\begin{enumerate}
    \item Order the node indices in $G$ so that $k_1\geq k_2\geq \cdots \geq k_N$.
    \item Initialize $V_h\leftarrow\{\}$, $h\leftarrow 0$, $M_h\leftarrow 0$, $h_{\text{ERs}}^\ast \leftarrow 0$, and set $\mathcal{L}_{\text{ERs}}^\ast\leftarrow \mathcal{L}^{(\text{ERs})}_0$ using Eq.~\ref{eq:L0ERs}.
    \item For $i\in \{1,...,N\}$:
    \begin{enumerate}
        \item Add $i$ to $V_h$ and update $h\leftarrow h+1$, $M_h\leftarrow M_h+k_i$.
        \item Compute the new description length $\mathcal{L}_h=\mathcal{L}^{(\text{ERs})}(V_h)$
        \item If $\mathcal{L}_h < \mathcal{L}_{\text{ERs}}^\ast$, set $\mathcal{L}_{\text{ERs}}^\ast\leftarrow \mathcal{L}_h$ and $h_{\text{ERs}}^\ast\leftarrow h$. Otherwise, do nothing.  
    \end{enumerate}
    \item After the loop terminates, the optimal set of hubs will be the set of node indices $\{1,...,h_{\text{ERs}}^\ast\}$, and these hubs will result in a description length of $\mathcal{L}_{\text{ERs}}^\ast$ in Eq.~\ref{eq:LERs}. 
\end{enumerate}

In the case of ties---i.e., if at a certain degree cutoff $k^\ast$ it is information theoretically optimal to only consider some fraction $0<f<1$ of the nodes $i$ with $k_i=k^\ast$ as hubs---we will check the cases $f=0$ (include all nodes with $k_i\geq k^\ast+1$) and $f=1$ (include all nodes with $k_i\geq k^\ast$) and choose the case with the lower description length. One can alternatively add all nodes with each unique degree $k$ at once during the greedy optimization process, to ensure that all nodes at or above the optimal threshold degree $k^\ast$ are included as hubs. These modifications remove the need to randomly break the tie to assign only some nodes of degree $k^\ast$ as hubs, since all such nodes are treated equivalently in our scheme. In principle, one can ignore this step and a random subset of nodes of degree $k^\ast$ will be be chosen as hubs based on the initial node ordering. This will often result in a slight compression gain at the cost of arbitrarily choosing the lowest-degree hubs.

The above method has an $\text{O}(N\log N)$ time complexity if the in-degrees $\bm{k}$ of the network $G$ are known, the bottleneck being sorting the degree sequence. Therefore it is equally as simple in practice to compute as hotspot identification methods such as the Loubar method and average degree method discussed in~\cite{louail2014mobile}. This method will also select for the number of hubs $h^\ast$ automatically based on the number that results in the best compression: too few hubs means we have not fully exploited the heterogeneity of the hub in-degrees for compression, and too many hubs means we have too little separation in the in-degrees of hubs and non-hubs to provide any meaningful compression. One key advantage of this approach over existing methods is that it allows for the result $h^\ast=0$ if there is no information theoretic justification to include any nodes as hubs according to Eq.~\ref{eq:LERs}. We will see in Sec.~\ref{sec:results} that this situation is quite common for networks with homogeneous in-degree distributions. 

There are a number of alternative ways we can construct an encoding that exploits a hub/non-hub dichotomy and allows for identifying an optimal set of network hubs. For example, one can transmit the hub nodes' incident edges individually using a two-step encoding inspired by the one used to compute Eq.~\ref{eq:L0CMs}, but transmit the remaining edges $E\setminus E_h$ using the same one-step encoding as in Eq.~\ref{eq:L0ERs}. This results in a description length of
\begin{align}\label{eq:LCMs}
\mathcal{L}^{(\text{CMs})}(V_h) &= \log NM+ \log {N\choose h} + \log{M_h+h-1\choose h-1} \\
&+ \sum_{i\in V_h}\log {N-1\choose k_i} + \log {(N-h)(N-1)\choose M-M_h} \nonumber.   
\end{align}
Here we also use the convention $\mathcal{L}^{(\text{CMs})}=\mathcal{L}^{(\text{CMs})}_0$ for $h=0,N$ for convenience.

In Appendix~\ref{appendix:opt-CM}, we show that this description length can be optimized over hub node sets $V_h$ using an analogous greedy algorithm, but in this case we only have a guarantee of local optimality. (One can in principle simply enforce the constraint that any hub node in-degree must be greater than or equal to any non-hub node in-degree, in which case this distinction is irrelevant since we will always add nodes in decreasing order of in-degree regardless of their effect on the description length.) 

After running the hub identification algorithm using either the ERs or CMs encoding, we can determine the extent to which a hub/non-hub dichotomy allows us to compress the network data $G$ in the first place---this gives us an alternative aggregate measure of heterogeneity in a network's in-degree distribution. To do this we compare the optimal level of compression $\mathcal{L}^\ast$ achieved with the hub-based encoding (either ERs or CMs) to the baseline compression levels in Eq.~\ref{eq:L0ERs} and Eq.~\ref{eq:L0CMs} when no hubs are utilized. The resulting quantity, which we call the inverse compression ratio, is given by
\begin{align}\label{eq:etas}
\eta^{(s)} = \frac{\mathcal{L}^\ast}{\text{max}(\mathcal{L}^{(\text{ERs})}_0,\mathcal{L}^{(\text{CMs})}_0)}.    
\end{align}
We can see that $\eta^{(s)} \in [0,1]$, since each encoding will give a minimum description length $\mathcal{L}^\ast$ that is at least as low as the description length for the encoding with no hub nodes. An inverse compression ratio near $0$ indicates that the hub nodes account for a large portion of the in-degrees and provide efficient compression of the network data, while an inverse compression ratio near $1$ indicates that the hubs do not contribute a significant portion of the in-degrees of the network. The case $\eta^{(s)}=1$ occurs when when we achieve no compression using hubs, which happens when the optimal number of hubs according to the encoding is $h^\ast=0$.

As done with Eq.~\ref{eq:L0ERs} and Eq.~\ref{eq:L0CMs}, we show in Appendix~\ref{appendix:ERvsCM} that the compression of the encoding corresponding to Eq.~\ref{eq:LCMs} is generally better than the encoding corresponding to Eq.~\ref{eq:LERs} in the high in-degree regime. In applications with networks $G$ that do not satisfy the conditions of Appendix~\ref{appendix:ERvsCM}, one can perform model selection among the two hub-based encodings (corresponding to Eq.s~\ref{eq:LERs} and \ref{eq:LCMs}) by identifying which description length is smaller. The more compressive encoding can then be used as the method for identifying the hub nodes in $G$. 

In addition to the in-degrees, one can consider using the \emph{out}-degrees to aid in network compression. This amounts to the same process except it identifies nodes with high out-degrees rather than in-degrees. This characterization is sensible, for example, in applications aiming to identify potential ``superspreaders'' in epidemic and misinformation modeling \cite{madotto2016super,grinberg2019fake}. We explore both the in- and out-degree versions of these methods in Sec.~\ref{sec:results}.

\subsection{Compressing directed multigraphs and weighted networks with hub nodes}
\label{sec:multigraph}

One can also extend the method discussed in Sec.~\ref{sec:simple} to directed multigraphs or directed, weighted networks with non-negative integer-valued weights. Such networks often arise in applications involving population ``flows'' from node to node, for example in transportation and human mobility modelling~\cite{verma2014revealing,roucolle2020measuring,mimar2022connecting,aguilar2022impact}, which were the original motivating examples for the Loubar method of \cite{louail2014mobile}. In this case, both the baseline encodings corresponding to Eq.s~\ref{eq:L0ERs} and~\ref{eq:L0CMs} as well as the hub-based encodings corresponding to Eq.s~\ref{eq:LERs} and~\ref{eq:LCMs} must be modified to account for the potential of having more than one edge (or, equivalently, an edge weight greater than $1$) between each pair of nodes. We will now let $M$ be the total weight of all the edges, and $k_i$ be the total weight of edges incident on node $i$---the latter is sometimes called the ``in-strength'' of node $i$, but we will continue to use the term ``in-degree'' for consistency with the simple graph case. We then have the same relation $M=\sum_{i=1}^{N}k_i$ as in the simple graph case.

The resulting description lengths, for which we use the superscripts ``m'' to denote ``multigraph'', are given by:
\begin{align}
\label{eq:L0ERm}
\mathcal{L}^{(\text{ERm})}_0 &= \log \multiset{N^2}{M},  \\
\label{eq:L0CMm}
\mathcal{L}^{(\text{CMm})}_0 &= \log {M+N-1\choose N-1} + \sum_{i=1}^{N}\log \multiset{N}{k_i}, \\
\label{eq:LERm}
\mathcal{L}^{(\text{ERm})}(V_h) &= \log NM + \log {N\choose h} + \log \multiset{hN}{M_h} \\
&~~~~~~~~~~~+ \log \multiset{(N-h)N}{M-M_h} \nonumber,\\
\label{eq:LCMm}
\mathcal{L}^{(\text{CMm})}(V_h) &= \log NM + \log {N\choose h} + \log{M_h+h-1\choose h-1} \\
&+ \sum_{i\in V_h}\log \multiset{N}{k_i} + \log \multiset{(N-h)N}{M-M_h}. \nonumber
\end{align}
These can be derived by transforming the binomial coefficients in Eq.s~\ref{eq:L0ERs},~\ref{eq:L0CMs},~\ref{eq:LERs}, and~\ref{eq:LCMs} to multiset coefficients $\multiset{n}{k}$, which count the number of ways to place $k$ edges into $n$ edge positions while allowing for repetitions. We also allow for self-edges, which requires the substitution $N-1\to N$ for the number of valid edge positions incident on a single node and $N(N-1)\to N^2$ for the total number of potential edge positions. We also use the conventions $\mathcal{L}^{(\text{ERm})}= \mathcal{L}^{(\text{ERm})}_0$ and $\mathcal{L}^{(\text{CMm})}= \mathcal{L}^{(\text{CMm})}_0$ for $h\in\{0,N\}$, as for the simple graph description lengths. 

Eq.s~\ref{eq:LERm} and~\ref{eq:LCMm} can be optimized using the same greedy procedure as in Sec.~\ref{sec:simple} for identifying the hubs from Eq.~\ref{eq:LERs} (with appropriate transformation of the variables). The resulting hub identification schemes can be used to construct an inverse compression ratio analogous to Eq.~\ref{eq:etas}, thus
\begin{align}
\label{eq:etam}
\eta^{(m)} = \frac{\mathcal{L}^\ast}{\text{max}(\mathcal{L}^{(\text{ERm})}_0,\mathcal{L}^{(\text{CMm})}_0)}, 
\end{align}
where $\mathcal{L}^\ast$ is the minimum description length achieved with the method of interest (ERm or CMm). Similar to the simple graph case, these ratios are bounded within $[0,1]$ and equal $1$ in the extreme case when $h^\ast=0$. $\mathcal{L}^\ast$ can be compared across the ERm and CMm encodings to select the more compressive encoding for a given multigraph $G$. 

In Fig.~\ref{fig:diagram} we show a schematic summarizing the hub-based encodings described in the last two sections. Code implementing the hub identification methods of this paper can be found at \url{https://github.com/aleckirkley/Network-hubs}.


\section{Results}
\label{sec:results}

\subsection{Hubs in networks with specified degree sequences}
\label{sec:synthetic}

We first perform a range of experiments with synthetic network data in order to understand the conditions under which hub nodes will be found using the methods described in Sec.~\ref{sec:methods}. We begin by analyzing the hub properties of random networks with degree sequences $\bm{k}$ that vary in magnitude and variability. Since the hubs and compression achieved using our proposed methods only depend on the degree sequence, we may interpret $\bm{k}$ in these experiments as the in-degree sequence of a network that is otherwise completely randomized, and we need not actually generate any network for each simulation.  

We choose three discrete probability distributions from which we generate the in-degrees $\bm{k}$: (1) A Poisson distribution, whose relative variance $\text{Variance}/\text{Mean}^2=1/\mu$ will vanish for large mean degree $\mu=\expec{k}$; (2) a Geometric distribution, whose relative variance $1-\mu^{-1}$ will tend to a constant for large mean degree; and a Power Law (e.g. Zipf) distribution, whose relative variance can potentially diverge. To ensure that all generated degree sequences $\bm{k}$ are graphically realizable, we consider the generated graphs to be multigraphs (or, equivalently, integer-weighted graphs) and use the encodings corresponding to Eq.~\ref{eq:LERm} and Eq.~\ref{eq:LCMm}.

\begin{table}[h]
\renewcommand{\arraystretch}{1.5}
\begin{tabular}{ | p{1.6cm}| p{6.9cm}| }
 \hline
\textbf{Name} & \textbf{Description}\\
\hline
        ER & Identifies hub nodes by iterating over $\bm{k}$ and greedily adding hubs to minimize the description length in Eq.~\ref{eq:LERs} (for simple graphs) or Eq.~\ref{eq:LERm} (for multigraphs and integer-weighted graphs). \\
        \hline
        CM & Identifies hub nodes by iterating over $\bm{k}$ and greedily adding hubs to minimize the description length in Eq.~\ref{eq:LCMs} (for simple graphs) or Eq.~\ref{eq:LCMm} (for multigraphs and integer-weighted graphs).\\
        \hline
        Average & Identifies hub nodes as all nodes $i\in V$ such that $k_i\geq \expec{k}=M/N$. \\
        \hline
        Loubar \cite{louail2014mobile} & Identifies hub nodes as the nodes at or above the $\big[1-\frac{\expec{k}}{\text{max}(\bm{k})}\big]$-th quantile in terms of degree. \\
        \hline
\end{tabular}
\caption{\textbf{Four methods for identifying hub nodes in Sec.~\ref{sec:results}.} All methods can be applied to in-degrees/out-degrees as well as unweighted/weighted networks by defining the degree sequence $\bm{k}$ accordingly.}
\label{tab:methods}
\end{table}

We compare our methods with two widely used methods for identifying hub nodes (``hotspots'') using the weighted in-degrees of human mobility networks \cite{louail2014mobile,louail2015uncovering,xu2019inverted,aguilar2022impact}. The first is the ``Average'' method, which simply classifies all nodes with in-degree values higher than the network average $M/N$ as hubs. However, the average method may not be conservative enough to give a useful guide for locating hotspots in human mobility applications, so the ``Loubar'' method is proposed in \cite{louail2014mobile}, which uses the Lorenz curve to construct a threshold for hub nodes that depends on the mean in-degree and the maximum in-degree. Our measures and these alternative measures are summarized in Table~\ref{tab:methods} for convenience. There is no absolute way in which one can decide which of these measures is ``best''---this may depend on the application of interest, and will require the consideration of multiple aspects of each measure---but our experiments highlight some potential intuitive advantages of the approaches based on the MDL principle.

\begin{figure*}
    \centering
    \includegraphics[width=\textwidth]{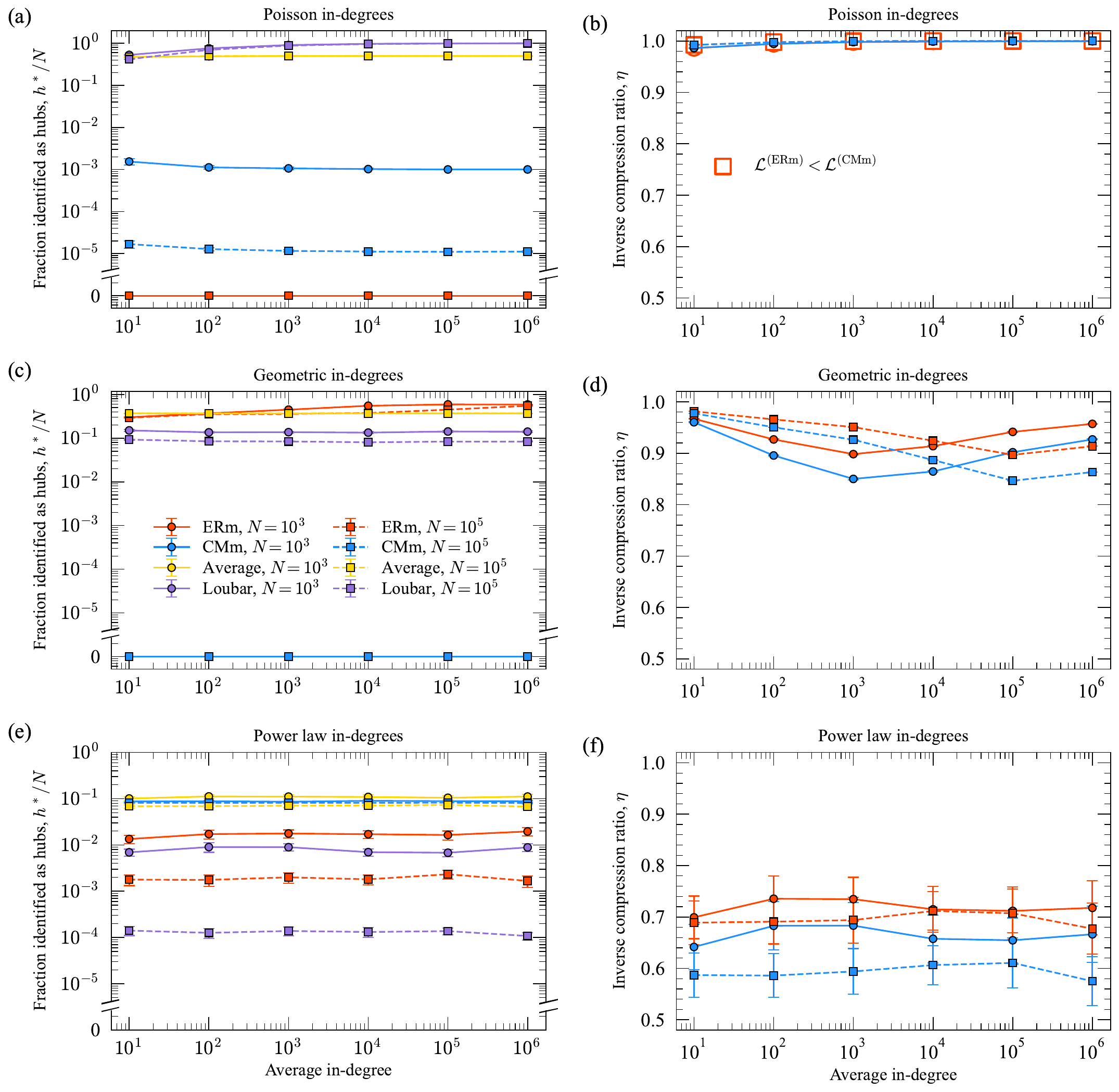}
    \caption{
    \textbf{Identifying hubs in networks with different in-degree distributions.} (a) Fraction of nodes $h^\ast/N$ identified as hubs using the four methods detailed in Table~\ref{tab:methods}, for Poisson distributed weighted in-degrees. Experiments were performed over a broad range of average in-degree for $N=10^3$ (solid lines) and $N=10^5$ (dotted lines). (b) Inverse compression ratio $\eta$ (Eq.~\ref{eq:etam}) for the ERm and CMm methods over the same set of experiments. The experiments were repeated for Geometrically distributed in-degrees (panels (c) and (d)) and Power Law (Zipf) distributed in-degrees (panels (e) and (f)), which exhibit progressively higher levels of relative variance. Error bars indicate two standard errors in the mean over 50 generated in-degree distributions, and large circles/squares around the data points in panel (b) indicate configurations for which the ERm model provided superior compression to the CMm model. 
    }
    \label{fig:distributions}
\end{figure*}

In Fig.~\ref{fig:distributions} we show the results of applying all four methods to degree sequences $\bm{k}$ randomly generated from the three distributions described above, for average in-degrees $\expec{k}\in \{10^1,10^2,10^3,10^4,10^5,10^6\}$ and network sizes $N\in \{10^3,10^5\}$. Each distribution only has a single parameter and can be specified uniquely given the desired mean in-degree. In each simulation, we generate $\bm{k}$ from the specified distribution, apply the four methods in Table~\ref{tab:methods}, and take the average result for the optimal number of hubs $h^\ast$ and inverse compression ratio (for the ERm and CMm encodings) over 50 repeated trials. For easier visualization, we plot $h^\ast /N$ to see what fraction of nodes are classified as hubs using each method. 

In Fig.~\ref{fig:distributions}a, we observe that for the Poisson distributed in-degrees, the ER and CM encodings both find very few hubs---the ER encoding always finds zero hubs, while the CM encoding only finds a handful. This is consistent with the small relative variance of the Poisson distribution, which will rarely result in any nodes having substantially larger in-degrees than the rest of the nodes. On the other hand, we can see that the Average and Loubar methods both indicate many hubs for networks with Poisson-distributed in-degrees. The Average method produces many hubs because the distribution is relatively symmetric, so a little less than half of the nodes will have in-degrees above the average. The Loubar method also produces many hubs because by construction it will indicate that the fraction of nodes that are hubs is $\expec{k}/\text{max}(\bm{k})$, and this quantity will be nearly equal to $1$ for Poisson samples with large mean in-degrees. We can see that in general the mean degree and network size do not play a particularly important role, as the fraction of hubs $h^\ast/N$ from each method is fairly constant across all simulation settings. The exception is the CM encoding, which produces only a handful (less than 10) hubs for most simulations, which results in smaller values of $h^\ast /N$ for larger $N$. 

In Fig.~\ref{fig:distributions}b, we can see that for Poisson distributed in-degrees, neither method (ERm or CMm) provides any substantial compression of the network, as the inverse compression ratios are approximately equal to $1$. This is because there is very little heterogeneity in the in-degrees that a hub-based information encoding can exploit to reduce the transmission cost of the data to a receiver. We also observe that the ER encoding has a slight edge over the CM encoding in terms of compression (indicated by the outlined markers), because transmitting the edges incident to hub nodes separately incurs an extra initial transmission cost but provides negligible additional compression. 

In Fig.~\ref{fig:distributions}c and Fig.~\ref{fig:distributions}d, we see a different story for Geometrically distributed in-degrees. In this case, the ER and Average methods indicate a substantial fraction of hub nodes, while the Loubar method is more conservative and only indicates that roughly $10\%$ or nodes are hubs. Meanwhile, the CM encoding is still very conservative, this time classifying zero nodes as hubs in all cases. Here the CM encoding now compresses better than the ER encoding, and both methods provide substantially better compression than in the Poisson case. This is consistent with the greater relative variance of the Geometric distribution, which will result in more effective compression using hub-based methods, and will benefit in particular from transmitting edges incident to the high degree hubs individually. There is also a greater dependence on the size $N$ of the network, and optimal compression for both encodings appears to be achieved at roughly $\expec{k}\approx N$. 

In Fig.~\ref{fig:distributions}e and Fig.~\ref{fig:distributions}f, we plot the same results for the Power Law in-degrees, which exhibit different behavior from the first two cases. Here we can see that the CM encoding is identifying a more substantial number of hubs (roughly $10\%$ of all nodes), while the ER encoding is a bit more conservative and the Loubar method is the most conservative. In this case one may expect a greater fraction of nodes to be identified as hubs than for Poisson or Geometric in-degrees, due to the highly skewed nature of the Power Law distribution, which would suggest that the CM results overall are the most consistent with our expectations. Both the ER and CM encodings are most compressive for Power Law in-degrees, reducing the information needed to transmit the network by roughly $30-40\%$, and the CM encoding becomes even more heavily favored in terms of inverse compression ratio. We also see a greater dependence on network size $N$ and higher sample variability in the Power Law results, as expected from the diverging relative variance in many cases. 

Overall, these experiments suggest that the CM encoding of Sec.~\ref{sec:methods} is performing the most consistent with intuition for these randomized degree sequences, as it finds zero or only a handful of hub nodes for Poisson and Geometric degree sequences, while finding that roughly $10\%$ of nodes are hubs for the Power Law degree sequences. Meanwhile, the ER method appears to be quite lenient for classifying hubs with Geometrically distributed in-degrees, but identifies a sensible number of hubs for the Poisson and Power Law cases---in particular, it identifies zero hubs for Poisson-distributed in-degrees. On the other hand, the Average and Loubar methods identify fewer hubs as the relative variance increases (from Poisson to Geometric to Power Law in-degrees).

\subsection{Emergence of hubs in growing networks}
\label{sec:growth}

Here we examine the four methods of Table~\ref{tab:methods} in a more dynamic context, applying these methods to track the evolution of hub nodes in growing networks. To simulate growing networks with varying levels of degree heterogeneity, we use the Price model for citation dynamics \cite{price1976general} with a variable attachment exponent. In this model, at each timestep $t=1,...,T$ a new node $i$ arrives with an out-degree $m$ and attaches each out-edge to an existing node $j$ with probability 
\begin{align}\label{eq:price}
q^{(t)}_{i\to j}=\frac{(k^{(t)}_j+1)^\alpha}{\sum_{j\in V^{(t)}}(k^{(t)}_j+1)^\alpha},    
\end{align}
where $k^{(t)}_j$ is the in-degree of node $j$ at time $t$ and $V^{(t)}$ are the nodes in the network at time $t$ (excluding $i$). The model begins with $m$ seed nodes of degree zero, which are the targets for the first incoming node's out-edges. In \cite{krapivsky2000connectivity} they find that a similar model with $m=1$ results in stretched exponential in-degree distributions for sublinear attachment ($\alpha < 0$) as $T\to\infty$, while for $\alpha >1$ the single initial node becomes a hub that has connections from nearly every other node that is added. Meanwhile, for $\alpha=1$ the in-degrees follow a Power Law asymptotically. These results suggest that as $\alpha\to 0$ we should see fewer and fewer hub nodes, while for $\alpha > 1$ we should see $h\approx m$ as $T\to\infty$ since  nearly every connection is made to a seed node. On the other hand, for $\alpha=1$ we expect to see a Power Law distribution of in-degrees and will observe a point in time at which a large number of hubs emerge.

\begin{figure*}
    \centering
    \includegraphics[width=\textwidth]{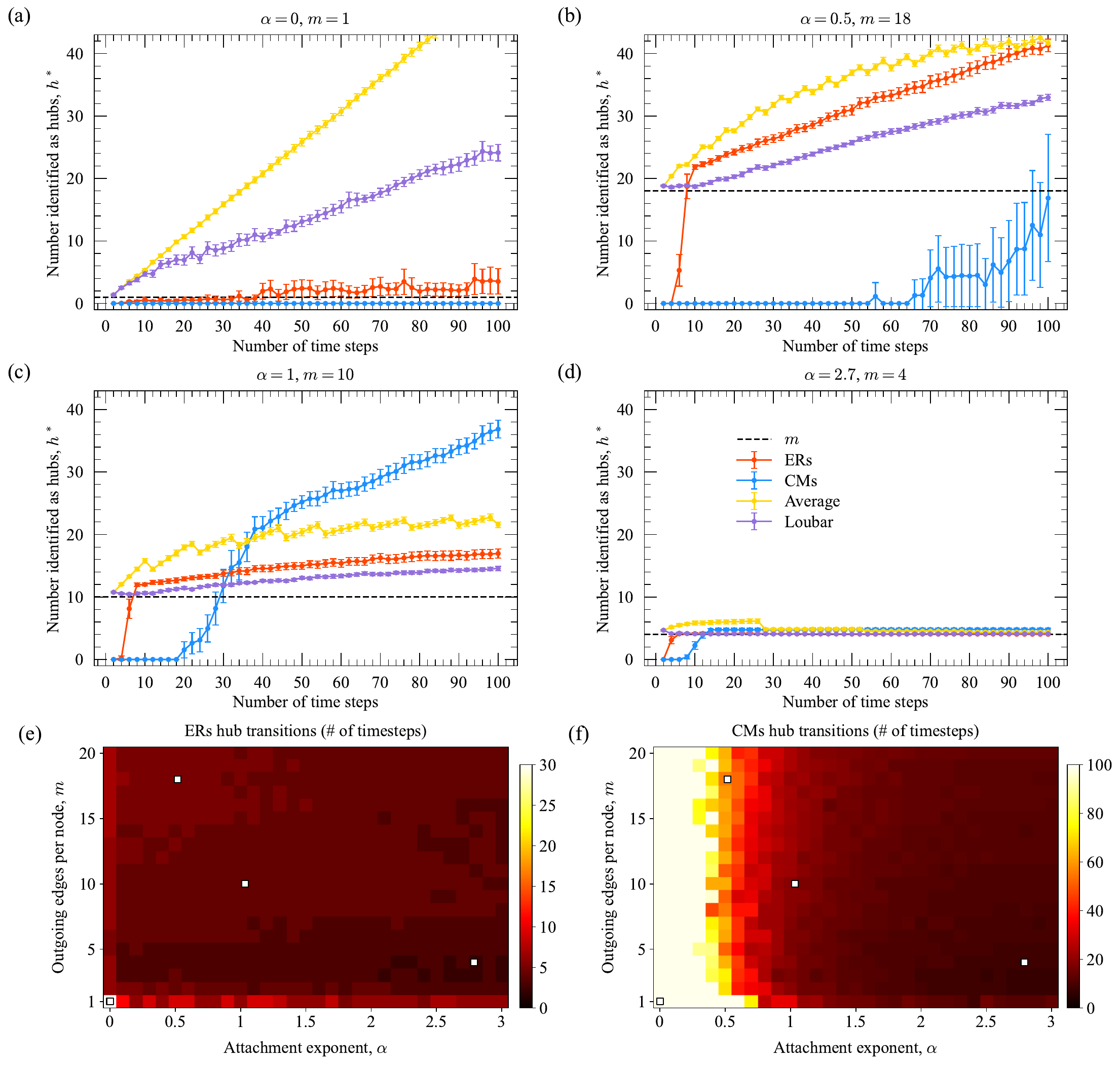}
    \caption{
    \textbf{Identifying hub transitions in Price's model with different attachment exponents and seed sets.} (a)-(d) The number of hubs $h^\ast$ identified by the four methods in Table~\ref{tab:methods} is shown as a function of the number of time steps in a generalization of Price's network growth model (Eq.~\ref{eq:price}) \cite{price1976general,krapivsky2000connectivity} for various attachment exponents $\alpha$ and numbers of seed nodes $m$ (dashed black lines). Error bars indicate two standard errors in the mean over 50 growth simulations with $T=100$ timesteps. (e)-(f) Expected number of timesteps until a single hub is detected (the ``hub transition'') over a range of attachment exponents and seed set sizes, for the ERs (Eq.~\ref{eq:LERs}) and CMs (Eq.~\ref{eq:LCMs}) hub identification objectives. Small white squares indicate the parameter values corresponding to panels (a)-(d). 
    }
    \label{fig:growth}
\end{figure*}

In Fig.~\ref{fig:growth}a, we show the average number of hubs $h^\ast$ detected by the four methods as a function of the number of time steps $t$, for 50 simulations of the growth model in Eq.~\ref{eq:price} with $T=100$ and $\{m,\alpha\}=\{1,0\}$. Under this parameter configuration, the network is a Random Recursive Tree, which will exhibit a highly homogeneous degree distribution and produce high degree hub nodes with very low probability \cite{janson2005asymptotic}. We can see that in this case, the ER and CM methods identify very few hubs---the CM method identifies exactly zero hubs in all simulations---while the Average and Loubar methods classify a sizable number of nodes as hubs (which increases steadily as the network grows). We don't see any sharp transition at which hubs emerge in the network according to any of the four methods. 

In Fig.~\ref{fig:growth}b, we show the growth model simulation results for the sublinear case of $\{m,\alpha\}=\{18,0.5\}$. We find again that the Average and Loubar methods find a steadily increasing number of hubs starting with $h^\ast\approx m$, with some small oscillations in $h^\ast$ for the Average method due to slight changes in the number of nodes with in-degree above the average of $m$. The ER and CM methods in this case exhibit quite different trends than for the Random Recursive Tree. The ER curve exhibits a sharp phase transition-like jump at $t\approx 5$, then steadily increases as the network grows further. This indicates that, in expectation, after roughly $5$ attachment events it is information theoretically more efficient to describe the network using hub nodes, according to Eq.~\ref{eq:LERs}. The CM encoding, typically being more conservative in its classification of hubs, does not begin to find hubs in the network until much farther into the simulations at around $t\approx 60$. By $T=100$, the CM method typically detects around $h^\ast\approx 18-20$ hubs, often finding that the seed nodes have high enough in-degrees to justify their existence as hubs under the CM encoding. 

In Fig.~\ref{fig:growth}c, we increase the attachment exponent so that linear preferential attachment will occur with $\{m,\alpha\}=\{10,1\}$, producing a Power Law in-degree distribution asymptotically. Here we see largely the same trend as in  Fig.~\ref{fig:growth}b for the ER, Average, and Loubar methods, with slower growth rates in $h^\ast$ as the simulations progress. This is consistent with a greater fraction of the in-connections being concentrated at the seed nodes, whose early existence in the network has given them a strong cumulative advantage. The CM encoding now classifies a substantial fraction of nodes as hubs, and displays sharp phase transition-like behavior at $t\approx 20$. This is consistent with the behavior seen in Fig.~\ref{fig:distributions}e for the Zipf-distributed in-degrees, where the CM encoding identified many hubs. 

In Fig.~\ref{fig:growth}d, we increase the attachment exponent to lie in the superlinear regime with $\{m,\alpha\}=\{4,2.7\}$. In this case, all curves converge to $h^\ast\approx m=4$, which reflects the fact that most incoming edges will attach to the $4$ seed nodes due to the superlinear attachment process. We again see the ER and CM encodings producing a sharp hub transition, but at even earlier time steps, while the Average and Loubar methods produce smooth curves as before.

Finally, in Fig.s~\ref{fig:growth}e and~\ref{fig:growth}f we plot the hub transition---the time step $t$ at which $h^\ast=1$ in expectation over the simulations---for the ER and CM encodings, at various values of the simulation parameters $\{m,\alpha\}$. The four parameter configurations corresponding to panels (a)-(d) are indicated by small white squares. We can see that for no attachment preference ($\alpha=0$) or a single outgoing edge ($m=1$), the ER model indicates a late hub transition, which we find is due to a smooth ascent reminiscent of that in Fig.~\ref{fig:growth}a (corresponding to the configuration in the bottom left corner of panel (e)). We find a weak trend in the ER hub transition as a function of the simulation parameters outside of these cases, with the hub transition occurring very early in the growth simulations. Meanwhile, the CM model has a hub transition that exhibits a much stronger dependence on the attachment exponent $\alpha$, with meaningful hub transitions occurring at roughly $\alpha=0.5$ for most values of $m$, against which the CM results display little variation  in panel (f).   

Altogether the results of these simulations further suggest that the ER and CM encodings---and particularly the CM encoding---are identifying meaningful hub structure in controlled synthetic network data that is consistent with expectations based on the evolution of the networks' in-degree sequences. In the next section, we explore the application of these methods to a corpus of real networks from various disciplines.

\subsection{Hubs in real-world networks}
\label{sec:real}

\begin{figure*}
    \centering
    \includegraphics[width=\textwidth]{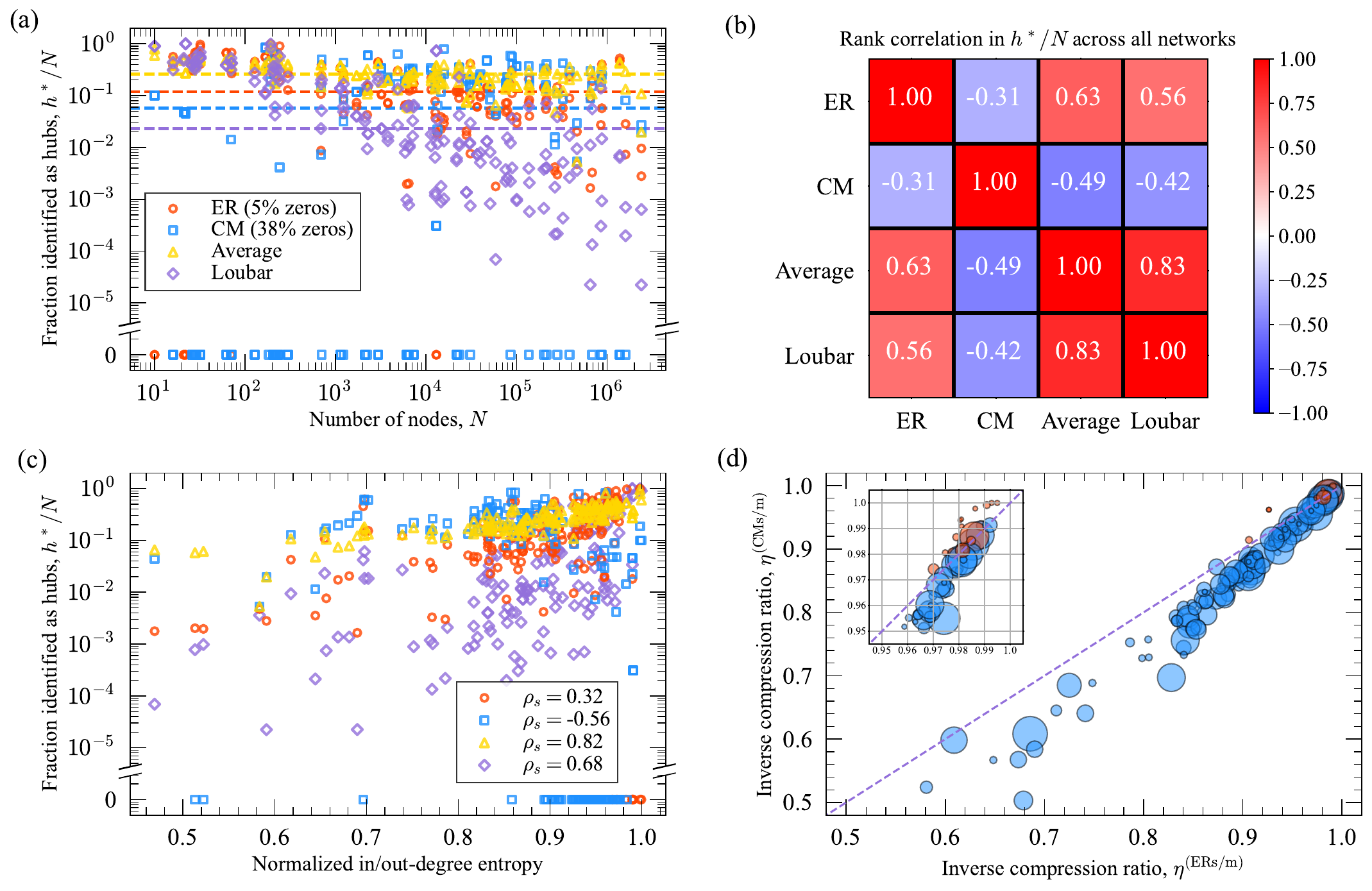}
    \caption{
    \textbf{Hub properties of real-world directed networks.} (a) The fraction $h^\ast/N$ of nodes identified as hubs using the four methods in Table~\ref{tab:methods}, for 82 real-world directed networks of various sizes collected from the Netzschleuder repository \cite{netzschleuder}. The median fraction of hubs found across all networks is shown with a dashed line for each method. See Appendix~\ref{appendix:networks} for details on the networks studied. (b) Spearman rank correlation in the fraction of nodes identified as hubs across all networks in the corpus, for each pair of methods examined. (c) Fraction of nodes identified as hubs vs the normalized degree entropy (Eq.~\ref{eq:entropy}) for the four methods. Spearman correlations between $h^\ast /N$ and the normalized degree entropy values are reported in the legend, and the marker colors/styles correspond to the methods indicated in panel (a). (d) Inverse compression ratios (Eq.~\ref{eq:etas} for simple graphs and Eq.~\ref{eq:etam} for weighted graphs) across all networks when using the ER and CM encodings (x- and y-axes respectively). The points are scaled monotonically with the size $N$ of the network analyzed, and red (blue) markers indicate that the ER (CM) encoding was more compressive for the given network. The inset shows a zoomed in view of the plot for $0.95\leq \eta\leq 1$, and the line of equality is shown as a dashed line for reference.  
    }
    \label{fig:real}
\end{figure*}

We collected 82 real-world network datasets from the Netzschleuder repository~\cite{netzschleuder} by querying all networks for which `is\_directed==True', `is\_bipartite==False', and for which the number of edges $M$ is less than $10^7$. Networks with non-integral weights were transformed to unweighted networks for the analyses, and after preprocessing there were 51 simple graphs---to which the ERs and CMs encodings were applied---and 31 weighted graphs/multigraphs---to which the ERm and CMm encodings were applied---for the analyses. The collected networks exhibit high variation in their size $N$ and average degree $M/N$, and represent systems from a broad range of disciplines (see Appendix~\ref{appendix:networks} for more details.)

In Fig.~\ref{fig:real}a, we plot the fractional number of hubs $h^\ast/N$ as a function of the number of nodes $N$ for all networks studied (both weighted and unweighted), using both the in- and out-degree distributions (giving 164 data points for each of the four methods in Table~\ref{tab:methods}). We find a story consistent with the findings for synthetic networks in Sec.~\ref{sec:synthetic}. In particular, the ER and CM encodings both assign zero hubs in a substantial fraction of cases, with the CM method assigning zero hubs in over a third (38\%) of the networks. In the cases where both methods detect hub nodes, the ER and CM methods detect a similar number of hubs, with the ER typically detecting slightly fewer. As seen in Fig.~\ref{fig:distributions}c, the Average method classifies a consistent \emph{fraction} of nodes $h^\ast/N\approx 0.3$ as hubs across the range of network sizes $N$. Meanwhile the Loubar method is much more conservative and classifies a consistent \emph{number} of nodes $h^\ast\approx 10$ as hubs as $N$ varies in Fig.~\ref{fig:real}a. The median values of $h^\ast/N$ for the four methods varies by roughly an order of magnitude, with $\text{Median}_{\text{Average}}\approx 0.26 > \text{Median}_{\text{ER}}\approx 0.12 > \text{Median}_{\text{CM}}\approx 0.057 >\text{Median}_{\text{Loubar}}\approx 0.023$. 

In Fig.~\ref{fig:real}b, we compute the Spearman correlation coefficient between the $h^\ast/N$ values produced by each pair of methods over all networks studied (in other words, the correlations in the y-values of panel (a).) We find that while the Average and Loubar methods are highly correlated with each other in terms of the fraction of hubs they identify across networks ($\rho\approx 0.83$), they are each more weakly correlated with the ER encoding's results ($\rho\approx 0.63$ and $\rho\approx 0.56$ respectively for the Average and Loubar methods). The CM method, meanwhile, is \emph{negatively} correlated with the other three methods in terms of $h^\ast/N$---for a given network, when the CM encoding classifies a greater fraction of hubs than usual, other methods will tend to classify a smaller fraction of hubs than usual. 

To investigate these correlations among the measures further we plot the fraction of hubs versus the normalized entropy $H_{\text{norm}}(\bm{k})$ of the degree sequence $\bm{k}$, given by
\begin{align}\label{eq:entropy}
H_{\text{norm}}(\bm{k}) = -\frac{1}{\log N}\sum_{i=1}^{N}\frac{k_i}{M}\log \left(\frac{k_i}{M}\right),    
\end{align}
for both in- and out-degrees $\bm{k}$ of each network in the corpus. The normalized degree entropy of Eq.~\ref{eq:entropy} is a natural measure of the variability in the degrees $\bm{k}$ which is bounded in $[0,1]$, with $H_{\text{norm}}=0$ being the extreme where all edges point to a single node and $H_{\text{norm}}=1$ being the other extreme where all nodes have identical degrees. We can see from Fig.~\ref{fig:real}c that the fraction of hubs $h^\ast/N$ for the Average and Loubar methods exhibits a fairly strong positive correlation with the normalized degree entropy of Eq.~\ref{eq:entropy}, while the ER encoding results are more weakly correlated with the degree entropy. In contrast, the CM encoding tends to assign a lower fraction of hub nodes as the degree entropy increases, as indicated by its strong negative correlation with $H_{\text{norm}}$. The behavior of the CM encoding is perhaps most aligned with expectations in that more heterogeneous degree sequences (those with lower entropy) should have more hubs, whereas degree sequences that are highly homogeneous (those with high entropy) should not have any hubs or only have relatively few hubs. This is also consistent with the observations in Fig.~\ref{fig:distributions}.

In Fig.~\ref{fig:real}d, we plot the inverse compression ratio (Eq.~\ref{eq:etas} for simple graphs and Eq.~\ref{eq:etam} for weighted graphs) for the ER and CM encodings (x- and y- axis respectively). We observe fairly substantial compression of the real networks using these hub-based encodings---more than $10\%$ of the information required for transmission is reduced relative to the baseline encoding in many cases---and that the compression achieved with each method is relatively independent of network size $N$. We can also see that the CM encoding outperforms the ER encoding in most cases, but that the ER encoding is more compressive for small networks and networks where little compression is achievable (i.e. networks with homogeneous degree sequences). This is consistent with the results of Appendix~\ref{appendix:ERvsCM}.


\section{Conclusion}

Here we described a set of methods for identifying hub nodes in directed networks with weighted or unweighted edges whose goal is to extract the subset of high degree nodes that allows for the best compression of the network data. Our methods are nonparametric, selecting the number of hub nodes in the network automatically using the Minimum Description Length principle, and can be run with a time complexity that is $\text{O}(N\log N)$ in the number of nodes $N$ in the network. We apply these methods in a range of experiments involving real and synthetic network data, finding an intuitive dependence on the degree heterogeneity of networks and improved performance relative to existing methods that are not explicitly designed for compressing network data. These methods provide a simple, principled, and flexible toolkit for exploring the hub structure of network data in a range of applications. 

There are a number of ways this work can be extended in future studies. Firstly, methods for compressing network data are not limited to focusing on purely local structure \cite{Peixoto14a,hebert2022network}, so one can in principle develop information theoretic encodings that exploit hub structure at larger scales in order to classify nodes that are more globally central in the network as hub nodes. However, the description length of such global encodings may become very challenging to compute due to the combinatorial structure of the problem. One can also adapt the ideas in this work to undirected graphs, which are more challenging to deal with than the directed graphs considered in this paper because edges only need to be specified in a single direction. In this case, one may aim to find a set of nodes that constitutes a complete or nearly complete vertex cover of the graph as the hub nodes that provide the most efficient network compression. One can also compare the compression of the methods proposed here with other methods such as stochastic blockmodels \cite{Peixoto2023Implicit} or various configuration models \cite{hebert2022network}, as well as integrate hub-based priors in these models for improved compression. Finally, one can apply the proposed methods to human mobility networks in order to uncover hotspot structure and compare empirical performance with the Average and Loubar methods in the mobility context \cite{louail2014mobile}.


\section*{Acknowledgments}
\vspace{-\baselineskip}
This research was supported by the HKU-100 Start Up Grant and the HKU Urban Systems Institute Fellowship Grant.


\begin{thebibliography}{10}
\expandafter\ifx\csname url\endcsname\relax
  \def\url#1{\texttt{#1}}\fi
\expandafter\ifx\csname urlprefix\endcsname\relax\def\urlprefix{URL }\fi

\bibitem{newman18c}
M.~Newman, \textit{Networks}. Oxford University Press, Oxford, 2nd edition (2018).

\bibitem{van2013network}
M.~P. Van~den Heuvel and O.~Sporns, Network hubs in the human brain. \textit{Trends in Cognitive Sciences} \textbf{17}(12), 683--696 (2013).

\bibitem{bassett2017network}
D.~S. Bassett and O.~Sporns, Network neuroscience. \textit{Nature Neuroscience} \textbf{20}(3), 353--364 (2017).

\bibitem{verma2014revealing}
T.~Verma, N.~A. Ara{\'u}jo, and H.~J. Herrmann, Revealing the structure of the world airline network. \textit{Scientific Reports} \textbf{4}(1), 5638 (2014).

\bibitem{roucolle2020measuring}
C.~Roucolle, T.~Seregina, and M.~Urdanoz, Measuring the development of airline networks: Comprehensive indicators. \textit{Transportation Research Part A: Policy and Practice} \textbf{133}, 303--324 (2020).

\bibitem{he2006hubs}
X.~He and J.~Zhang, Why do hubs tend to be essential in protein networks? \textit{PLoS Genetics} \textbf{2}(6), e88 (2006).

\bibitem{mimar2022connecting}
S.~Mimar, D.~Soriano-Pa{\~n}os, A.~Kirkley, H.~Barbosa, A.~Sadilek, A.~Arenas, J.~G{\'o}mez-Garde{\~n}es, and G.~Ghoshal, Connecting intercity mobility with urban welfare. \textit{PNAS Nexus} \textbf{1}(4), pgac178 (2022).

\bibitem{aguilar2022impact}
J.~Aguilar, A.~Bassolas, G.~Ghoshal, S.~Hazarie, A.~Kirkley, M.~Mazzoli, S.~Meloni, S.~Mimar, V.~Nicosia, J.~J. Ramasco, \textit{et~al.}, Impact of urban structure on infectious disease spreading. \textit{Scientific Reports} \textbf{12}(1), 3816 (2022).

\bibitem{Barabasi16}
A.-L. Barab\'asi, \textit{Network Science}. Cambridge University Press, Cambridge (2016).

\bibitem{chakrabarti2008epidemic}
D.~Chakrabarti, Y.~Wang, C.~Wang, J.~Leskovec, and C.~Faloutsos, Epidemic thresholds in real networks. \textit{ACM Transactions on Information and System Security (TISSEC)} \textbf{10}(4), 1--26 (2008).

\bibitem{gradon2021countering}
K.~T. Grado{\'n}, J.~A. Ho{\l}yst, W.~R. Moy, J.~Sienkiewicz, and K.~Suchecki, Countering misinformation: A multidisciplinary approach. \textit{Big Data \& Society} \textbf{8}(1), 20539517211013848 (2021).

\bibitem{kleinberg1999authoritative}
J.~M. Kleinberg, Authoritative sources in a hyperlinked environment. \textit{Journal of the ACM (JACM)} \textbf{46}(5), 604--632 (1999).

\bibitem{oldham2019consistency}
S.~Oldham, B.~Fulcher, L.~Parkes, A.~Arnatkevici{\=u}t{\.e}, C.~Suo, and A.~Fornito, Consistency and differences between centrality measures across distinct classes of networks. \textit{PloS One} \textbf{14}(7), e0220061 (2019).

\bibitem{ronqui2015analyzing}
J.~R.~F. Ronqui and G.~Travieso, Analyzing complex networks through correlations in centrality measurements. \textit{Journal of Statistical Mechanics: Theory and Experiment} \textbf{2015}(5), P05030 (2015).

\bibitem{grando2016analysis}
F.~Grando, D.~Noble, and L.~C. Lamb, An analysis of centrality measures for complex and social networks. In \textit{2016 IEEE Global Communications Conference (GLOBECOM)}, pp. 1--6, IEEE (2016).

\bibitem{schoch2017correlations}
D.~Schoch, T.~W. Valente, and U.~Brandes, Correlations among centrality indices and a class of uniquely ranked graphs. \textit{Social Networks} \textbf{50}, 46--54 (2017).

\bibitem{bloch2023centrality}
F.~Bloch, M.~O. Jackson, and P.~Tebaldi, Centrality measures in networks. \textit{Social Choice and Welfare} pp. 1--41 (2023).

\bibitem{shao2018rank}
C.~Shao, P.~Cui, P.~Xun, Y.~Peng, and X.~Jiang, Rank correlation between centrality metrics in complex networks: an empirical study. \textit{Open Physics} \textbf{16}(1), 1009--1023 (2018).

\bibitem{wu2011information}
S.~Wu and S.~Wang, Information-theoretic outlier detection for large-scale categorical data. \textit{IEEE Transactions on Knowledge and Data Engineering} \textbf{25}(3), 589--602 (2011).

\bibitem{akoglu2015graph}
L.~Akoglu, H.~Tong, and D.~Koutra, Graph based anomaly detection and description: a survey. \textit{Data Mining and Knowledge Discovery} \textbf{29}, 626--688 (2015).

\bibitem{bohm2009coco}
C.~B{\"o}hm, K.~Haegler, N.~S. M{\"u}ller, and C.~Plant, Coco: coding cost for parameter-free outlier detection. In \textit{Proceedings of the 15th ACM SIGKDD International Conference on Knowledge Discovery and Data Mining}, pp. 149--158 (2009).

\bibitem{louail2014mobile}
T.~Louail, M.~Lenormand, O.~G. Cantu~Ros, M.~Picornell, R.~Herranz, E.~Frias-Martinez, J.~J. Ramasco, and M.~Barthelemy, From mobile phone data to the spatial structure of cities. \textit{Scientific Reports} \textbf{4}(1), 5276 (2014).

\bibitem{hazarie2021interplay}
S.~Hazarie, D.~Soriano-Pa{\~n}os, A.~Arenas, J.~G{\'o}mez-Garde{\~n}es, and G.~Ghoshal, Interplay between population density and mobility in determining the spread of epidemics in cities. \textit{Communications Physics} \textbf{4}(1), 191 (2021).

\bibitem{louail2015uncovering}
T.~Louail, M.~Lenormand, M.~Picornell, O.~Garcia~Cantu, R.~Herranz, E.~Frias-Martinez, J.~J. Ramasco, and M.~Barthelemy, Uncovering the spatial structure of mobility networks. \textit{Nature Communications} \textbf{6}(1), 6007 (2015).

\bibitem{xu2019inverted}
W.~Xu, H.~Chen, E.~Frias-Martinez, M.~Cebrian, and X.~Li, The inverted u-shaped effect of urban hotspots spatial compactness on urban economic growth. \textit{Royal Society Open Science} \textbf{6}(11), 181640 (2019).

\bibitem{rissanen1978}
J.~Rissanen, Modeling by the shortest data description. \textit{Automatica} \textbf{14}, 465--471 (1978).

\bibitem{peixoto2019bayesian}
T.~P. Peixoto, Bayesian stochastic blockmodeling. \textit{Advances in Network Clustering and Blockmodeling} pp. 289--332 (2019).

\bibitem{hebert2022network}
L.~H{\'e}bert-Dufresne, J.-G. Young, A.~Daniels, and A.~Allard, Network onion divergence: Network representation and comparison using nested configuration models with fixed connectivity, correlation and centrality patterns. \textit{arXiv preprint arXiv:2204.08444}  (2022).

\bibitem{gallagher2021clarified}
R.~J. Gallagher, J.-G. Young, and B.~F. Welles, A clarified typology of core-periphery structure in networks. \textit{Science Advances} \textbf{7}(12), eabc9800 (2021).

\bibitem{kirkley2022spatial}
A.~Kirkley, Spatial regionalization based on optimal information compression. \textit{Communications Physics} \textbf{5}(1), 249 (2022).

\bibitem{kirkley2023compressing}
A.~Kirkley, A.~Rojas, M.~Rosvall, and J.-G. Young, Compressing network populations with modal networks reveal structural diversity. \textit{Communications Physics} \textbf{6}(1), 148 (2023).

\bibitem{kirkley2023constructing}
A.~Kirkley, Constructing hypergraphs from temporal data. \textit{arXiv preprint arXiv:2308.16546}  (2023).

\bibitem{karrer2011stochastic}
B.~Karrer and M.~E. Newman, Stochastic blockmodels and community structure in networks. \textit{Physical Review E} \textbf{83}(1), 016107 (2011).

\bibitem{madotto2016super}
A.~Madotto and J.~Liu, Super-spreader identification using meta-centrality. \textit{Scientific Reports} \textbf{6}(1), 38994 (2016).

\bibitem{grinberg2019fake}
N.~Grinberg, K.~Joseph, L.~Friedland, B.~Swire-Thompson, and D.~Lazer, Fake news on twitter during the 2016 us presidential election. \textit{Science} \textbf{363}(6425), 374--378 (2019).

\bibitem{price1976general}
D.~d.~S. Price, A general theory of bibliometric and other cumulative advantage processes. \textit{Journal of the American Society for Information Science} \textbf{27}(5), 292--306 (1976).

\bibitem{krapivsky2000connectivity}
P.~L. Krapivsky, S.~Redner, and F.~Leyvraz, Connectivity of growing random networks. \textit{Physical Review Letters} \textbf{85}(21), 4629 (2000).

\bibitem{janson2005asymptotic}
S.~Janson, Asymptotic degree distribution in random recursive trees. \textit{Random Structures \& Algorithms} \textbf{26}(1-2), 69--83 (2005).

\bibitem{netzschleuder}
T.~P. Peixoto, The {N}etzschleuder network catalogue and repository (2020). Accessible at \url{https://networks.skewed. de}.

\bibitem{Peixoto14a}
T.~P. Peixoto, Hierarchical block structures and high-resolution model selection in large networks. \textit{Physical Review X} \textbf{4}, 011047 (2014).

\bibitem{Peixoto2023Implicit}
T.~P. Peixoto and A.~Kirkley, Implicit models, latent compression, intrinsic biases, and cheap lunches in community detection. \textit{Physical Review E} \textbf{108}, 024309 (2023).

\end{thebibliography}


\clearpage
\appendix
\onecolumngrid

\section{Relative compression of $\mathcal{L}^{(\text{ERs})}_0$ and $\mathcal{L}^{(\text{CMs})}_0$}
\label{appendix:relative0ercm}

Before proceeding, we can establish the useful inequality:
\begin{align}\label{eq:identity}
\log {\sum_n x_n \choose \sum_n y_n} - \sum_n\log {x_n\choose y_n} \geq 0
\end{align}
for non-negative integers $\{x_n\},\{y_n\}$. Letting $Y=\sum_n y_n$, the Vandermonde identity gives
\begin{align}
{\sum_n x_n \choose Y} &= \sum_{\sum_nz_n=Y} \prod_{n}{x_n\choose z_n} 
\geq \prod_{n}{x_n\choose y_n},
\end{align}
and taking the logarithm of both sides of the inequality gives the desired result.

Applying Stirling's approximation to the first term in Eq.~\ref{eq:L0CMs} gives, for $N\gg 1$, the following expression
\begin{align}
\log {M+N-1\choose N-1} &\approx (M+N-1)H_b\left (\frac{N-1}{M+N-1}\right)\\
&=[N(\expec{k}+1)-1]H_b\left (\frac{N\expec{k}}{N(\expec{k}+1)-1}\right)\\
&\approx N(\expec{k}+1)H_b\left(\frac{\expec{k}}{\expec{k}+1}\right)\\
&\approx N\log (\expec{k}+1),
\end{align}
where $H_b(p)=-p\log p -(1-p)\log (1-p)$ is the binary entropy function. We can then see that for $N\gg \expec{k}\gg 1$, this term will vanish relative to the second term in Eq.~\ref{eq:L0CMs}, and so we have
\begin{align}
\mathcal{L}^{(\text{CMs})}_0 \approx \sum_{i=1}^{N}\log {N-1\choose k_i}.    
\end{align}
Now, applying the identity in Eq.~\ref{eq:identity} we can see that
\begin{align}
\mathcal{L}^{(\text{ERs})}_0 - \mathcal{L}^{(\text{CMs})}_0 \approx \log {N(N-1) \choose M} - \sum_{i=1}^{N}\log {N-1\choose k_i} \geq 0.  
\end{align}
Therefore, in the regime $N\gg \expec{k} \gg 1$, we will always achieve superior compression using the two-step encoding with description length in  Eq.~\ref{eq:L0CMs} than the one-step encoding used for Eq.~\ref{eq:L0ERs}. However, for small and/or very sparse networks we do not have any guarantee that $\mathcal{L}^{(\text{CMs})}_0 \leq \mathcal{L}^{(\text{ERs})}_0$ since the above approximation is no longer valid. In this regime, the one-step encoding may compress better than the two-step encoding since the initial transmission cost for the degrees is non-negligible.

\section{Optimization of $\mathcal{L}^{(\text{ERs})}(V_h)$ and $\mathcal{L}^{(\text{ERm})}(V_h)$}
\label{appendix:opt-ER}

We can show that the global optimum of $\mathcal{L}^{(\text{ERs})}(V_h)$ is obtained using the greedy hub identification process outlined in Sec.~\ref{sec:simple}. This demonstrates that the optimal set of $h$ hub nodes is the set of $h$ nodes with the highest in-degrees. For any fixed $h\geq 1$, we have that substituting $M_h\to M_h+1$---in other words, an increase in the cumulative in-degree $M_h$ of the hubs (see Eq.~\ref{eq:LERs})---induces a change $\Delta_h^{(\text{ERs})}$ in the description length $\mathcal{L}^{(\text{ERs})}$ of
\begin{align}\label{eq:deltaERs}
\Delta_h^{(\text{ERs})} &= \log \frac{[h(N-1)-M_h][M-M_h]}{[M_h+1][(N-1)(N-h)-(M-M_h)+1]}\\
&< \log \frac{[h(N-1)-M_h][M-M_h]}{M_h[(N-1)(N-h)-(M-M_h)]}\\
&=\log \frac{h(N-1)M-h(N-1)M_h-MM_h+M_h^2}{M_hN(N-1)-h(N-1)M_h-MM_h+M_h^2}.
\end{align}
Now, as long as $h(N-1)M\leq M_hN(N-1)$---equivalent to the condition $M_h/h \geq (M/N)$, or that the average degree of the hubs is greater than or equal to the average in-degree of the network as a whole---then we have that $\Delta_h^{(\text{ERs})} < 0$, which in turn implies that the optimal set of $h$ nodes to choose as hubs are those with the $h$ highest in-degrees, since this set of nodes will maximize $M_h$. Therefore, if we force the first hub node (i.e. for $h=1$) to be the node of maximum in-degree, then the greedy scheme from Sec.~\ref{sec:methods} must produce a globally optimal set of hubs with respect to the description length in Eq.~\ref{eq:LERs}. 

Similarly, we can show that the global optimum of $\mathcal{L}^{(\text{ERm})}(V_h)$ is also obtained using the greedy hub identification process outlined in Sec.~\ref{sec:simple} (with quantities appropriately mapped from the ERs encoding to the ERm encoding). The analogous expression to Eq.~\ref{eq:deltaERs} for the multigraph description length in Eq.~\ref{eq:LERm} is
\begin{align}
\Delta_h^{(\text{ERm})} &= \log \frac{[M_h+hN][M-M_h]}{[M_h+1][M-M_h+N(N-h)-1]}\\
&=\log \frac{[MM_h-M_h^2]+hN[M-M_h]}{[MM_h-M_h^2]+[M-M_h]+N(N-h)[M_h+1]-[M_h+1]}\\
&<\log \frac{[MM_h-M_h^2]+hN[M-M_h]}{[MM_h-M_h^2]+[N(N-h)-1][M_h+1]}.
\end{align}
In this case, we need $hN[M-M_h] \leq [N(N-h)-1][M_h+1]$ to hold in order for $\Delta_h^{(\text{ERm})}<0$ (and, hence, global optimality of the greedy method). Rearranging the inequality, we can see that for $M_h,N(N-h)\gg 1$, the same condition $M_h/h>M/N$ will guarantee $\Delta_h^{(\text{ERm})}<0$. Therefore, forcing the highest in-degree node as the hub for $h=1$ allows for global optimality of the greedy hub identification algorithm in this case as well, so long as the maximum in-degree is much greater than $1$.

\section{Optimization of $\mathcal{L}^{(\text{CMs})}(V_h)$ and $\mathcal{L}^{(\text{CMm})}(V_h)$}
\label{appendix:opt-CM}

We can guarantee local optimality of the greedy algorithm for $\mathcal{L}^{(\text{CMs})}(V_h)$ by showing that the description length resulting from adding a node of degree $k+1$ at step $h$ is always less than the description length resulting from adding a node of degree $k$ at step $h$. In other words, at any step $h$, we should add the remaining non-hub node with the highest in-degree as the new hub. The difference $\Delta^{(\text{CMs})}_{k\vert h}$ in Eq.~\ref{eq:LCMs} due to adding a node of degree $k+1$ at step $h$ instead of a node of degree $k$ at step $h$ is given by
\begin{align}
\Delta^{(\text{CMs})}_{k\vert h} = \log \frac{[M_{h-1}+h+k][N-k-1][M-M_{h-1}-k]}{[M_{h-1}+k+1][k+1][(N-h)(N-1)-(M-M_{h-1})+k+1]},    
\end{align}
where here we've let $M_{h-1}$ be the cumulative in-degree of whatever set of hubs we've chosen after step $h-1$. Setting $h/N\to\gamma$, $M/N\to \expec{k}$, and $M_{h-1}/(h-1)\to \expec{k}_{h-1}$, we have that in the limit $N\gg 1$ the expression can be simplified to
\begin{align}
\Delta^{(\text{CMs})}_{k\vert h}\approx \log \frac{[\expec{k}_{h-1}+1][\expec{k}-\gamma \expec{k}_{h-1}]}{[k+1][\expec{k}_{h-1}-\gamma \expec{k}_{h-1}]}.    
\end{align}
From here, we can see that $\Delta^{(\text{CMs})}_{k\vert h}\leq 0$ as long as 
\begin{align}
\expec{k}_{h-1}[k-\expec{k}]+\gamma\expec{k}_{h-1}[\expec{k}_{h-1}-k]+[\expec{k}_{h-1}-\expec{k}]\geq 0,
\end{align}
which is satisfied for $\expec{k}_{h-1}\geq k\geq \expec{k}$. The first inequality is satisfied for the greedy scheme, since nodes are added in order of decreasing in-degree. Therefore, we have a guarantee of local optimality under the greedy scheme when $k\geq \expec{k}$. Repeating the above argument for $\mathcal{L}^{(\text{CMm})}(V_h)$ gives the same final condition $\expec{k}_{h-1}\geq k\geq \expec{k}$ for local optimality using the CMm encoding. Outside of this regime---i.e., when the node being considered has a degree $k$ that is less than the network average $\expec{k}$---we no longer have a proof of local optimality for the greedy scheme. However, as discussed in Sec.~\ref{sec:methods} we can simply enforce the constraint that every hub node must have a degree that is at least as large as the highest non-hub node degree, in which case the greedy scheme is the only scheme that will produce hub sets $V_h$ consistent with this constraint at all values of $h$. 

\section{Relative compression of ``ER'' vs ``CM'' hub-based encodings}
\label{appendix:ERvsCM}

In the regime $\expec{k}_h\equiv M_h/h\gg 1$, we have that 
\begin{align}
\log {M_h+h-1\choose h-1} &\approx [h(\expec{k}_h+1)-1]H_b\left(\frac{h\expec{k}_h}{h(\expec{k}_h+1)-1}\right) \approx h(\expec{k}_h+1)H_b\left(\frac{\expec{k}_h}{\expec{k}_h+1}\right)
\approx h\log(\expec{k}_h+1)
\end{align}
Now, using a similar argument to the one in Appendix~\ref{appendix:relative0ercm}, we have that this term will vanish relative to $\sum_{i\in V_h}\log {N-1\choose k_i}$, and subtracting Eq.~\ref{eq:LCMs} from Eq.~\ref{eq:LERs} gives
\begin{align}
\mathcal{L}^{(\text{ERs})}(V_h) - \mathcal{L}^{(\text{CMs})}(V_h)  
&=\log {h(N-1)\choose M_h} - \log {M_h+h-1\choose h-1} - \sum_{i\in V_h}\log {N-1\choose k_i}\\
&\approx \log {h(N-1)\choose M_h} - \sum_{i\in V_h}\log {N-1\choose k_i}\\
&\geq 0,
\end{align}
where we've used the identity in Eq.~\ref{eq:identity}. Therefore, in the regime $\expec{k}_h \gg 1$, we will always achieve superior compression using the ``CMs'' encoding over the ``ERs'' encoding. However, for very sparse networks or networks with no hub nodes, we do not have any guarantee that $\mathcal{L}^{(\text{CMs})} \leq \mathcal{L}^{(\text{ERs})}$ since the above approximation is no longer valid. In this regime, the ERs encoding may compress better than the CMs encoding since the transmission cost for the degrees is non-negligible. The same argument can be used to establish that in the same regime we will achieve superior compression using the CMm encoding over the ERm encoding.

\newpage
\section{Real-world network details}
\label{appendix:networks}

\centering
\begin{tabular}{|c|c|c|c|c|}
 \hline
\textbf{\#} &                 ~~~~~~~~~~~~~~\textbf{Name} \cite{netzschleuder}~~~~~~~~~~~~~~ &        ~~~~~~~\textbf{$N$}~~~~~~~ &              ~~~~~~~\textbf{$M$}~~~~~~~ &  ~~~~~~~\textbf{Weighted}~~~~~~~ \\ 
 \hline
0  &        packet\_delays &       10 &             45 &          True \\ 
 \hline
1  &        rhesus\_monkey &       16 &            120 &          True \\ 
 \hline
2  &    high\_tech\_company &       21 &            210 &          True \\ 
 \hline
3  &          moreno\_taro &       22 &            231 &         False \\ 
 \hline
4  &                bison &       26 &            325 &          True \\ 
 \hline
5  &         moreno\_sheep &       28 &            378 &          True \\ 
 \hline
6  &               cattle &       28 &            378 &          True \\ 
 \hline
7  &          7th\_graders &       29 &            406 &          True \\ 
 \hline
8  &                 hens &       32 &            496 &         False \\ 
 \hline
9  &     college\_freshmen &       32 &            496 &          True \\ 
 \hline
10 &             macaques &       62 &           1891 &          True \\ 
 \hline
11 &           highschool &       70 &           2415 &          True \\ 
 \hline
12 &             law\_firm &       71 &           2485 &          True \\ 
 \hline
13 &       foodweb\_baywet &      128 &           8128 &         False \\ 
 \hline
14 &        email\_company &      167 &          13861 &          True \\ 
 \hline
15 &  foodweb\_little\_rock &      183 &          16653 &         False \\ 
 \hline
16 &                  psi &      192 &          18336 &          True \\ 
 \hline
17 &        cintestinalis &      205 &          20910 &         False \\ 
 \hline
18 &            fao\_trade &      214 &          22791 &          True \\ 
 \hline
19 &       residence\_hall &      217 &          23436 &          True \\ 
 \hline
20 &        un\_migrations &      232 &          26796 &         False \\ 
 \hline
21 &      physician\_trust &      241 &          28920 &         False \\ 
 \hline
22 &       celegansneural &      297 &          43956 &         False \\ 
 \hline
23 &  yeast\_transcription &      690 &         295283 &          True \\ 
 \hline
24 &         messal\_shale &      700 &         244650 &         False \\ 
 \hline
25 &            uni\_email &     1133 &         641278 &         False \\ 
 \hline
26 &             polblogs &     1224 &         934396 &          True \\ 
 \hline
27 &           faa\_routes &     1226 &         750925 &         False \\ 
 \hline
28 &   interactome\_stelzl &     1706 &        1454365 &         False \\ 
 \hline
29 &        at\_migrations &     2115 &        2235555 &          True \\ 
 \hline
30 &   interactome\_figeys &     2239 &        2505441 &         False \\ 
 \hline
31 &       us\_air\_traffic &     2278 &        2593503 &          True \\ 
 \hline
32 &            fly\_larva &     2952 &        4355676 &         False \\ 
 \hline
33 &          openflights &     3214 &        5163291 &         False \\ 
 \hline
34 &        bitcoin\_alpha &     3783 &        7153653 &         False \\ 
 \hline
35 &            fediverse &     4860 &       11807370 &         False \\ 
 \hline
36 &        bitcoin\_trust &     5881 &       17290140 &         False \\ 
 \hline
37 &                 jung &     6120 &       18724140 &         False \\ 
 \hline
38 &                  jdk &     6434 &       20694961 &         False \\ 
 \hline
39 &             advogato &     6539 &       21379008 &          True \\ 
 \hline
40 &                 elec &     7118 &       25329403 &         False \\ 
 \hline
41 &                chess &     7301 &       26648650 &         False \\ 
 \hline
42 &             wiki\_rfa &    11381 &       64757890 &         False \\ 
 \hline
43 &            dblp\_cite &    12590 &       79247755 &         False \\ 
 \hline
44 &              anybeat &    12645 &       79941690 &         False \\ 
 \hline
45 &         chicago\_road &    12979 &       84224356 &          True \\ 
 \hline
\end{tabular}

\centering
\begin{tabular}{|c|c|c|c|c|}
 \hline
\textbf{\#} &                 ~~~~~~~~~~~~~~\textbf{Name} \cite{netzschleuder}~~~~~~~~~~~~~~ &        ~~~~~~~\textbf{$N$}~~~~~~~ &              ~~~~~~~\textbf{$M$}~~~~~~~ &  ~~~~~~~\textbf{Weighted}~~~~~~~ \\ 
 \hline
46 &               foldoc &    13356 &       89184690 &          True \\ 
 \hline
47 &              inploid &    14629 &      106996506 &         False \\ 
 \hline
48 &               google &    15763 &      124228203 &         False \\ 
 \hline
49 &        fly\_hemibrain &    21739 &      236281191 &          True \\ 
 \hline
50 &           word\_assoc &    23132 &      267533146 &          True \\ 
 \hline
51 &                 cora &    23166 &      268320195 &         False \\ 
 \hline
52 &           lkml\_reply &    27927 &      840498910 &          True \\ 
 \hline
53 &           digg\_reply &    30398 &      462004003 &         False \\ 
 \hline
54 &                linux &    30837 &      475444866 &         False \\ 
 \hline
55 &          email\_enron &    36692 &      673133086 &         False \\ 
 \hline
56 &           pgp\_strong &    39796 &      791840910 &         False \\ 
 \hline
57 &        facebook\_wall &    46952 &     1102221676 &         False \\ 
 \hline
58 &     slashdot\_threads &    51083 &     1304710903 &         False \\ 
 \hline
59 &    python\_dependency &    58743 &     1725340653 &         False \\ 
 \hline
60 &       epinions\_trust &    75879 &     2879167673 &          True \\ 
 \hline
61 &         slashdot\_zoo &    79116 &     3129687706 &          True \\ 
 \hline
62 &          twitter\_15m &    85712 &     3741488617 &          True \\ 
 \hline
63 &              prosper &    89269 &     3984432546 &         False \\ 
 \hline
64 &        wiki\_link\_dyn &   100312 &     5031198516 &         False \\ 
 \hline
65 &        lastfm\_aminer &   136409 &     9303639436 &         False \\ 
 \hline
66 &           wiki\_users &   138592 &     9603801936 &         False \\ 
 \hline
67 &         academia\_edu &   200169 &    20033714196 &         False \\ 
 \hline
68 &          google\_plus &   211187 &    22299868891 &         False \\ 
 \hline
69 &        flickr\_aminer &   214626 &    23032052625 &         False \\ 
 \hline
70 &             email\_eu &   265214 &    35169100291 &         False \\ 
 \hline
71 &         stanford\_web &   281903 &    39734791656 &          True \\ 
 \hline
72 &       notre\_dame\_web &   325729 &    53049527856 &         False \\ 
 \hline
73 &             citeseer &   384413 &    73886485078 &         False \\ 
 \hline
74 &              twitter &   465017 &   108120172636 &         False \\ 
 \hline
75 &            yahoo\_ads &   653260 &   213373987170 &         False \\ 
 \hline
76 &         berkstan\_web &   685230 &   234770419065 &          True \\ 
 \hline
77 &       myspace\_aminer &   854498 &   365082988753 &         False \\ 
 \hline
78 &           google\_web &   875713 &   402754552837 &          True \\ 
 \hline
79 &             wikitree &  1382751 &   955999472625 &         False \\ 
 \hline
80 &             trec\_web &  1601787 &  1282859995791 &         False \\ 
 \hline
81 &    wikipedia-en-talk &  2394385 &  2866538566920 &         False \\ 
 \hline
\end{tabular}

\end{document}